\documentclass[letterpaper,compsoc,twoside]{IEEEtran}
\usepackage{fixltx2e} % LaTeX patches, \textsubscript
\usepackage{cmap} % fix search and cut-and-paste in Acrobat
\usepackage{ifthen}
\usepackage[T1]{fontenc}
\usepackage[utf8]{inputenc}
\usepackage{amsmath}

\usepackage[font={small,it},labelfont=bf]{caption}
\usepackage{float}

\usepackage{textcomp} % text symbol macros

\pdfoutput=1
\usepackage{scipy}
\makeatletter
\def\PY@reset{\let\PY@it=\relax \let\PY@bf=\relax%
    \let\PY@ul=\relax \let\PY@tc=\relax%
    \let\PY@bc=\relax \let\PY@ff=\relax}
\def\PY@tok#1{\csname PY@tok@#1\endcsname}
\def\PY@toks#1+{\ifx\relax#1\empty\else%
    \PY@tok{#1}\expandafter\PY@toks\fi}
\def\PY@do#1{\PY@bc{\PY@tc{\PY@ul{%
    \PY@it{\PY@bf{\PY@ff{#1}}}}}}}
\def\PY#1#2{\PY@reset\PY@toks#1+\relax+\PY@do{#2}}

\expandafter\def\csname PY@tok@gd\endcsname{\def\PY@tc##1{\textcolor[rgb]{0.63,0.00,0.00}{##1}}}
\expandafter\def\csname PY@tok@gu\endcsname{\let\PY@bf=\textbf\def\PY@tc##1{\textcolor[rgb]{0.50,0.00,0.50}{##1}}}
\expandafter\def\csname PY@tok@gt\endcsname{\def\PY@tc##1{\textcolor[rgb]{0.00,0.27,0.87}{##1}}}
\expandafter\def\csname PY@tok@gs\endcsname{\let\PY@bf=\textbf}
\expandafter\def\csname PY@tok@gr\endcsname{\def\PY@tc##1{\textcolor[rgb]{1.00,0.00,0.00}{##1}}}
\expandafter\def\csname PY@tok@cm\endcsname{\let\PY@it=\textit\def\PY@tc##1{\textcolor[rgb]{0.25,0.50,0.56}{##1}}}
\expandafter\def\csname PY@tok@vg\endcsname{\def\PY@tc##1{\textcolor[rgb]{0.73,0.38,0.84}{##1}}}
\expandafter\def\csname PY@tok@m\endcsname{\def\PY@tc##1{\textcolor[rgb]{0.13,0.50,0.31}{##1}}}
\expandafter\def\csname PY@tok@mh\endcsname{\def\PY@tc##1{\textcolor[rgb]{0.13,0.50,0.31}{##1}}}
\expandafter\def\csname PY@tok@cs\endcsname{\def\PY@tc##1{\textcolor[rgb]{0.25,0.50,0.56}{##1}}\def\PY@bc##1{\setlength{\fboxsep}{0pt}\colorbox[rgb]{1.00,0.94,0.94}{\strut ##1}}}
\expandafter\def\csname PY@tok@ge\endcsname{\let\PY@it=\textit}
\expandafter\def\csname PY@tok@vc\endcsname{\def\PY@tc##1{\textcolor[rgb]{0.73,0.38,0.84}{##1}}}
\expandafter\def\csname PY@tok@il\endcsname{\def\PY@tc##1{\textcolor[rgb]{0.13,0.50,0.31}{##1}}}
\expandafter\def\csname PY@tok@go\endcsname{\def\PY@tc##1{\textcolor[rgb]{0.20,0.20,0.20}{##1}}}
\expandafter\def\csname PY@tok@cp\endcsname{\def\PY@tc##1{\textcolor[rgb]{0.00,0.44,0.13}{##1}}}
\expandafter\def\csname PY@tok@gi\endcsname{\def\PY@tc##1{\textcolor[rgb]{0.00,0.63,0.00}{##1}}}
\expandafter\def\csname PY@tok@gh\endcsname{\let\PY@bf=\textbf\def\PY@tc##1{\textcolor[rgb]{0.00,0.00,0.50}{##1}}}
\expandafter\def\csname PY@tok@ni\endcsname{\let\PY@bf=\textbf\def\PY@tc##1{\textcolor[rgb]{0.84,0.33,0.22}{##1}}}
\expandafter\def\csname PY@tok@nl\endcsname{\let\PY@bf=\textbf\def\PY@tc##1{\textcolor[rgb]{0.00,0.13,0.44}{##1}}}
\expandafter\def\csname PY@tok@nn\endcsname{\let\PY@bf=\textbf\def\PY@tc##1{\textcolor[rgb]{0.05,0.52,0.71}{##1}}}
\expandafter\def\csname PY@tok@no\endcsname{\def\PY@tc##1{\textcolor[rgb]{0.38,0.68,0.84}{##1}}}
\expandafter\def\csname PY@tok@na\endcsname{\def\PY@tc##1{\textcolor[rgb]{0.25,0.44,0.63}{##1}}}
\expandafter\def\csname PY@tok@nb\endcsname{\def\PY@tc##1{\textcolor[rgb]{0.00,0.44,0.13}{##1}}}
\expandafter\def\csname PY@tok@nc\endcsname{\let\PY@bf=\textbf\def\PY@tc##1{\textcolor[rgb]{0.05,0.52,0.71}{##1}}}
\expandafter\def\csname PY@tok@nd\endcsname{\let\PY@bf=\textbf\def\PY@tc##1{\textcolor[rgb]{0.33,0.33,0.33}{##1}}}
\expandafter\def\csname PY@tok@ne\endcsname{\def\PY@tc##1{\textcolor[rgb]{0.00,0.44,0.13}{##1}}}
\expandafter\def\csname PY@tok@nf\endcsname{\def\PY@tc##1{\textcolor[rgb]{0.02,0.16,0.49}{##1}}}
\expandafter\def\csname PY@tok@si\endcsname{\let\PY@it=\textit\def\PY@tc##1{\textcolor[rgb]{0.44,0.63,0.82}{##1}}}
\expandafter\def\csname PY@tok@s2\endcsname{\def\PY@tc##1{\textcolor[rgb]{0.25,0.44,0.63}{##1}}}
\expandafter\def\csname PY@tok@vi\endcsname{\def\PY@tc##1{\textcolor[rgb]{0.73,0.38,0.84}{##1}}}
\expandafter\def\csname PY@tok@nt\endcsname{\let\PY@bf=\textbf\def\PY@tc##1{\textcolor[rgb]{0.02,0.16,0.45}{##1}}}
\expandafter\def\csname PY@tok@nv\endcsname{\def\PY@tc##1{\textcolor[rgb]{0.73,0.38,0.84}{##1}}}
\expandafter\def\csname PY@tok@s1\endcsname{\def\PY@tc##1{\textcolor[rgb]{0.25,0.44,0.63}{##1}}}
\expandafter\def\csname PY@tok@gp\endcsname{\let\PY@bf=\textbf\def\PY@tc##1{\textcolor[rgb]{0.78,0.36,0.04}{##1}}}
\expandafter\def\csname PY@tok@sh\endcsname{\def\PY@tc##1{\textcolor[rgb]{0.25,0.44,0.63}{##1}}}
\expandafter\def\csname PY@tok@ow\endcsname{\let\PY@bf=\textbf\def\PY@tc##1{\textcolor[rgb]{0.00,0.44,0.13}{##1}}}
\expandafter\def\csname PY@tok@sx\endcsname{\def\PY@tc##1{\textcolor[rgb]{0.78,0.36,0.04}{##1}}}
\expandafter\def\csname PY@tok@bp\endcsname{\def\PY@tc##1{\textcolor[rgb]{0.00,0.44,0.13}{##1}}}
\expandafter\def\csname PY@tok@c1\endcsname{\let\PY@it=\textit\def\PY@tc##1{\textcolor[rgb]{0.25,0.50,0.56}{##1}}}
\expandafter\def\csname PY@tok@kc\endcsname{\let\PY@bf=\textbf\def\PY@tc##1{\textcolor[rgb]{0.00,0.44,0.13}{##1}}}
\expandafter\def\csname PY@tok@c\endcsname{\let\PY@it=\textit\def\PY@tc##1{\textcolor[rgb]{0.25,0.50,0.56}{##1}}}
\expandafter\def\csname PY@tok@mf\endcsname{\def\PY@tc##1{\textcolor[rgb]{0.13,0.50,0.31}{##1}}}
\expandafter\def\csname PY@tok@err\endcsname{\def\PY@bc##1{\setlength{\fboxsep}{0pt}\fcolorbox[rgb]{1.00,0.00,0.00}{1,1,1}{\strut ##1}}}
\expandafter\def\csname PY@tok@kd\endcsname{\let\PY@bf=\textbf\def\PY@tc##1{\textcolor[rgb]{0.00,0.44,0.13}{##1}}}
\expandafter\def\csname PY@tok@ss\endcsname{\def\PY@tc##1{\textcolor[rgb]{0.32,0.47,0.09}{##1}}}
\expandafter\def\csname PY@tok@sr\endcsname{\def\PY@tc##1{\textcolor[rgb]{0.14,0.33,0.53}{##1}}}
\expandafter\def\csname PY@tok@mo\endcsname{\def\PY@tc##1{\textcolor[rgb]{0.13,0.50,0.31}{##1}}}
\expandafter\def\csname PY@tok@mi\endcsname{\def\PY@tc##1{\textcolor[rgb]{0.13,0.50,0.31}{##1}}}
\expandafter\def\csname PY@tok@kn\endcsname{\let\PY@bf=\textbf\def\PY@tc##1{\textcolor[rgb]{0.00,0.44,0.13}{##1}}}
\expandafter\def\csname PY@tok@o\endcsname{\def\PY@tc##1{\textcolor[rgb]{0.40,0.40,0.40}{##1}}}
\expandafter\def\csname PY@tok@kr\endcsname{\let\PY@bf=\textbf\def\PY@tc##1{\textcolor[rgb]{0.00,0.44,0.13}{##1}}}
\expandafter\def\csname PY@tok@s\endcsname{\def\PY@tc##1{\textcolor[rgb]{0.25,0.44,0.63}{##1}}}
\expandafter\def\csname PY@tok@kp\endcsname{\def\PY@tc##1{\textcolor[rgb]{0.00,0.44,0.13}{##1}}}
\expandafter\def\csname PY@tok@w\endcsname{\def\PY@tc##1{\textcolor[rgb]{0.73,0.73,0.73}{##1}}}
\expandafter\def\csname PY@tok@kt\endcsname{\def\PY@tc##1{\textcolor[rgb]{0.56,0.13,0.00}{##1}}}
\expandafter\def\csname PY@tok@sc\endcsname{\def\PY@tc##1{\textcolor[rgb]{0.25,0.44,0.63}{##1}}}
\expandafter\def\csname PY@tok@sb\endcsname{\def\PY@tc##1{\textcolor[rgb]{0.25,0.44,0.63}{##1}}}
\expandafter\def\csname PY@tok@k\endcsname{\let\PY@bf=\textbf\def\PY@tc##1{\textcolor[rgb]{0.00,0.44,0.13}{##1}}}
\expandafter\def\csname PY@tok@se\endcsname{\let\PY@bf=\textbf\def\PY@tc##1{\textcolor[rgb]{0.25,0.44,0.63}{##1}}}
\expandafter\def\csname PY@tok@sd\endcsname{\let\PY@it=\textit\def\PY@tc##1{\textcolor[rgb]{0.25,0.44,0.63}{##1}}}

\def\PYZus{\char`\_}

\def\PYZgt{\char`\>}

\def\PYZdl{\char`\$}

\def\PYZsq{\char`\'}

\makeatother

\providecommand*{\DUfootnotemark}[3]{%
  \raisebox{1em}{\hypertarget{#1}{}}%
  \hyperlink{#2}{\textsuperscript{#3}}%
}
\providecommand{\DUfootnotetext}[4]{%
  \begingroup%
  \renewcommand{\thefootnote}{%
    \protect\raisebox{1em}{\protect\hypertarget{#1}{}}%
    \protect\hyperlink{#2}{#3}}%
  \footnotetext{#4}%
  \endgroup%
}

\providecommand*{\DUroletitlereference}[1]{\textsl{#1}}

\ifthenelse{\isundefined{\hypersetup}}{
  \usepackage[colorlinks=true,linkcolor=blue,urlcolor=blue]{hyperref}
  \urlstyle{same} % normal text font (alternatives: tt, rm, sf)
}{}

\begin{document}
\newcounter{footnotecounter}\title{Bloscpack: a compressed lightweight serialization format for numerical data}\author{Valentin Haenel$^{\setcounter{footnotecounter}{1}\fnsymbol{footnotecounter}\setcounter{footnotecounter}{2}\fnsymbol{footnotecounter}}$%
          \setcounter{footnotecounter}{1}\thanks{\fnsymbol{footnotecounter} %
          Corresponding author: \protect\href{mailto:valentin@haenel.co}{valentin@haenel.co}}\setcounter{footnotecounter}{2}\thanks{\fnsymbol{footnotecounter} Independent}\thanks{%

          \noindent%
          Copyright\,\copyright\,2014 Valentin Haenel. This is an open-access article distributed under the terms of the Creative Commons Attribution License, which permits unrestricted use, distribution, and reproduction in any medium, provided the original author and source are credited. http://creativecommons.org/licenses/by/3.0/%
        }}\maketitle
          \renewcommand{\leftmark}{PROC. OF THE 6th EUR. CONF. ON PYTHON IN SCIENCE (EUROSCIPY 2013)}
          \renewcommand{\rightmark}{BLOSCPACK: A COMPRESSED LIGHTWEIGHT SERIALIZATION FORMAT FOR NUMERICAL DATA}

\setcounter{page}{3}
\newcommand*{\docutilsroleref}{\ref}
\newcommand*{\docutilsrolelabel}{\label}
\AtEndDocument{\cleardoublepage}
\begin{abstract}This paper introduces the Bloscpack file format and the accompanying Python
reference implementation. Bloscpack is a lightweight, compressed binary
file-format based on the Blosc codec and is designed for lightweight, fast
serialization of numerical data. This article presents the features of the
file-format and some some API aspects of the reference implementation, in
particular the ability to handle Numpy ndarrays.  Furthermore, in order to
demonstrate its utility, the format is compared both feature- and
performance-wise to a few alternative lightweight serialization solutions
for Numpy ndarrays.  The performance comparisons take the form of some
comprehensive benchmarks over a range of different artificial datasets with
varying size and complexity, the results of which are presented as the last
section of this article.\end{abstract}

\section{Introduction%
  \label{introduction}%
}

When using compression during storage of numerical data there are two potential
improvements one can make. First, by using compression, naturally one can save
storage space. Secondly—and this is often overlooked—one can save time.
When using compression during serialization, the total compression time is the
sum of the time taken to perform the compression and the time taken to write
the compressed data to the storage medium. Depending on the compression speed
and the compression ratio, this sum maybe less than the time taken to serialize
the data in uncompressed format i.e.  $write_{uncompressed} >
write_{compressed} + time_{compress}$

The Bloscpack file format and Python reference implementation aims to
achieve exactly this by leveraging the fast, multithreaded, blocking and
shuffling Blosc codec.

\section{Blosc%
  \label{blosc}%
}

Blosc \cite{Blosc} is a fast, multitreaded, blocking and shuffling
compressor designed initially for in-memory compression. Contrary to
many other available compressors which operate sequentially on a data
buffer, Blosc uses the blocking technique \cite{Alted2009,Alted2010} to
split the dataset into individual blocks. It can then operate on each
block using a different thread which effectively leads to a
multithreaded compressor.  The block size is chosen such that it
either fits into a typical L1 cache (for compression levels up to 6)
or L2 cache (for compression levels larger than 6). In modern CPUs L1
and L2 are typically non-shared between other cores, and so this
choice of block size leads to an optimal performance during
multi-thread operation.

Also, Blosc features a shuffle filter \cite{Alted2009} (p.71) which may
reshuffle multi-byte elements, e.g. 8 byte doubles, by
significance. The net result for series of numerical elements with
little difference between elements that are close, is that similar
bytes are placed closer together and can thus be better compressed
(this is specially true on time series datasets). Internally, Blosc
uses its own codec, \emph{blosclz}, which is a derivative of FastLZ \cite{FastLZ}
and implements the LZ77 \cite{LZ77} scheme.  The reason for Blosc to
introduce its own codec is mainly the desire for simplicity (blosclz
is a highly streamlined version of FastLZ), as well as providing a
better interaction with Blosc infrastructure.

Moreover, Blosc is designed to be extensible, and allows other codecs than
blosclz to be used in it. In other words, one can consider Blosc as a
meta-compressor, in that it handles the splitting of the data into blocks,
optionally applying the shuffle filter (or other future filters), while being
responsible of coordinating the individual threads during operation. Blosc then
relies on a \textquotedbl{}real\textquotedbl{} codec to perform that actual compression of the data blocks.
As such, one can think of Blosc as a way to parallelize existing codecs, while
allowing to apply filters (also called pre-conditioners). In fact, at the time
when the research presented in this paper was conducted (Summer 2013),
a proof-of-concept implementation existed to integrate
the well known Snappy codec \cite{Snappy} as well as LZ4 \cite{LZ4} into the Blosc framework.
As of January 2014 this proof of concept has matured and as of version 1.3.0
Blosc comes equipped with support for Snappy \cite{Snappy}, LZ4 \cite{LZ4} and even Zlib {[}\hyperref[zlib]{zlib}{]}.

Blosc was initially developed to support in-memory compression in order to
mitigate the effects of the memory hierarchy \cite{Jacob2009}. More specifically,
to mitigate the effects of memory latency, i.e. the ever growing divide between
the CPU speed and the memory access speed–which is also known as the problem of
the \emph{starving CPUs} \cite{Alted2009}.

The goal of in-memory compression techniques is to have a numerical container
which keeps all data as in-memory compressed blocks. If the data needs to be
operated on, it is decompressed only in the caches of the CPU.
Hence, data can be moved faster from memory to CPU and the net result is faster
computation, since less CPU cycles are wasted while waiting for data. Similar
techniques are applied successfully in other settings. Imagine for example, one
wishes to transfer binary files over the internet. In this case the transfer
time can be significantly improved by compressing the data before transferring
it and decompressing it after having received it. As a result the total
compressed transfer time, which is taken to be the sum of the compression and
decompression process and the time taken to transfer the compressed file, is
less than the time taken to transfer the plain file. For example the well known
UNIX tool \texttt{rsync} \cite{rsync} implements a \DUroletitlereference{-z} switch which performs
compression of the data before sending it and decompression after receiving it.
The same basic principle applies to in-memory compression, except that we are
transferring data from memory to CPU.  Initial implementations based on Blosc
exist, c.f. Blaze \cite{Blaze} and carray \cite{CArray}, and have been shown to yield
favourable results {[}Personal communication with Francesc Alted{]}.

\section{Numpy%
  \label{numpy}%
}

The Numpy \cite{VanDerWalt2011,Numpy} ndarray is the de-facto multidimensional
numerical container for scientific python applications.  It is probably the
most fundamental package of the scientific python ecosystem and widely used and
relied upon by third-party libraries and applications. It consists of the
N-dimensional array class, various different initialization routines and many
different ways to operate on the data efficiently.

\section{Existing Lightweight Solutions%
  \label{existing-lightweight-solutions}%
}

There are a number of other plain (uncompressed) and compressed lightweight
serialization formats for Numpy arrays that we can compare Bloscpack to. We
specifically ignore more heavyweight solutions, such as HDF5, in this comparison.%
\begin{itemize}

\item 

NPY
\item 

NPZ
\item 

ZFile
\end{itemize}

\subsection{NPY%
  \label{npy}%
}

\emph{NPY} \cite{NPY} is a simple plain serialization format for numpy. It is considered
somewhat of a gold standard for the serialization. One of its advantages is
that it is very, very lightweight. The format specification is simple and can
easily be digested within an hour. In essence it simply contains the ndarray
metadata and the serialized data block. The metadata amounts to the \texttt{dtype}, the
\texttt{order} and the \texttt{shape} or the array. The main drawback is that it is a
plain serialization format and does not support compression.

\subsection{NPZ%
  \label{npz}%
}

\emph{NPZ} is, simply put, a Zip file which contains multiple NPY files. Since this is
a Zip file it may be optionally compressed, however the main uses case is to
store multiple ndarrays in a single file. Zip is an implementation of the
DEFLATE \cite{DEFLATE} algorithm. Unlike the other evaluated compressed formats,
NPZ does not support a compression level setting.

\subsection{ZFile%
  \label{zfile}%
}

\emph{ZFile} is the native serialization format that ships with the Joblib
\cite{Joblib} framework. Joblib is equipped with a caching mechanism that supports caching
input and output arguments to functions and can thus avoid running heavy
computations if the input has not changed. When serializing ndarrays with
Joblib, a special subclass of the Pickler is used to store the metadata whereas
the datablock is serialized as a ZFile. ZFile uses zlib \cite{zlib} internally and
simply runs zlib on the entire data buffer. zlib is also an implementation of
the DEFLATE algorithm. One drawback of the current ZFile implementation is that
no chunking scheme is employed. This means that the memory requirements might
be twice that of the original input. Imagine trying to compress an
incompressible buffer of 1GB: in this case the memory requirement would be 2GB,
since the entire buffer must be copied in memory as part of the compression
process before it can be written out to disk.

\section{Bloscpack Format%
  \label{bloscpack-format}%
}

The Bloscpack format and reference implementation builds a serialization format
around the Blosc codec. It is a simple chunked file-format well suited for the
storage of numerical data. As described in the Bloscpack format description,
the big-picture of the file-format is as follows:%
\begin{quote}\begin{verbatim}
|-header-|-meta-|-offsets-|

|-chunk-|-checksum-|-chunk-|-checksum-|...|
\end{verbatim}

\end{quote}
The format contains a 32 byte \texttt{header} which contains various options and
settings for the file, for example a magic string, the format version number
and the total number of chunks. The \texttt{meta} section is of variable size and
can contain any metadata that needs to be saved alongside the data.  An
optional \texttt{offsets} section is provided to allow for partial decompression of
the file in the future. This is followed by a series of \texttt{chunks}, each of
which is a blosc compressed buffer. Each chunk can be optionally followed by a
\texttt{checksum} of the compressed data which can help to protect against silent
data corruption.

The chunked format was initially chosen to circumvent a 2GB limitation of the
Blosc codec. In fact, the ZFile format suffers from this exact limitation
since zlib—at least the Python bindings—is also limited to buffers of
2GB in size. The limitation stems from the fact that \texttt{int32} are used
internally by the algorithms to store the size of the buffer and the maximum
value of an \texttt{int32} is indeed 2GB. In any case, using a chunked scheme turned
out to be useful in its own right. Using a modest chunk-size of e.g. 1MB (the
current default) causes less stress on the memory subsystem. This also means
that in contrast to ZFile, only a small fixed overhead equal to the chunk-size
is required during the compression and decompression process, for example when
compressing or decompression from/to an external storage medium.

With version 3 the format was enhanced to allow appending data to an existing
Bloscpack compressed file. This is achieved by over-allocating the offsets and
metadata section with dummy values to allow chunks to be appended later and
metadata to be enlarged. One caveat of this is that we can not pre-allocate an
infinite amount of space and so only a limited amount of data can potentially be
appended. However, to provide potential consumers of the format with as much
flexibility as possible, the amount of space to be pre-allocated is
configurable.

For an in-depth discussion of the technical details of the  Bloscpack format
the interested reader is advised to consult the official documentation
\cite{Bloscpack}. This contains a full description of the header layout, the
sizes of the entries and their permissible values.

\section{Command Line Interface%
  \label{command-line-interface}%
}

Initially, Bloscpack was conceived as a command-line compression tool. At the
time of writing, a Python API is in development and, in fact, the command-line
interface is being used to drive and dog-food the Python API. Contrary to
existing tools such as \texttt{gzip} \cite{gzip}, \texttt{bloscpack} doesn't use command-line
options to control its mode of operation, but instead uses the \emph{subcommand}
style. Here is a simple example:\begin{Verbatim}[commandchars=\\\{\},fontsize=\footnotesize]
\PY{g+gp}{\PYZdl{}} ./blpk compress data.dat
\PY{g+gp}{\PYZdl{}} ./blpk decompress data.dat.blp data.dcmp
\end{Verbatim}
Another interesting subcommand is \texttt{info} which can be used to inspect the
header and metadata of an existing file:\begin{Verbatim}[commandchars=\\\{\},fontsize=\footnotesize]
\PY{g+gp}{\PYZdl{}} ./blpk info data.dat.blp
\PY{g+go}{[...]}
\end{Verbatim}
The Bloscpack documentation contains extensive descriptions of the various
options and many examples of how to use the command line API.

\section{Packing Numpy Arrays%
  \label{packing-numpy-arrays}%
}

As of version 0.4.0 Bloscpack comes with support for serializing Numpy
ndarrays. The approach is simple and lightweight: the data buffer is saved in
Blosc compressed chunks as defined by the Bloscpack format. The \texttt{shape},
\texttt{dtype} and \texttt{order} attributes—the same ones saved in the NPY format—are saved in the metadata section.  Upon de-serialization, first an empty
ndarray is allocated from the information in the three metadata attributes.
Then, the Bloscpack chunks are decompressed directly into the pre-allocated
array.

The Bloscpack Python API for Numpy ndarray is very similar to the simple NPY
interface; arrays can be serialized/de-serialized using single function
invocations.

Here is an example of serializing a Numpy array to file:\begin{Verbatim}[commandchars=\\\{\},fontsize=\footnotesize]
\PY{g+gp}{\PYZgt{}\PYZgt{}\PYZgt{} }\PY{k+kn}{import} \PY{n+nn}{numpy} \PY{k+kn}{as} \PY{n+nn}{np}
\PY{g+gp}{\PYZgt{}\PYZgt{}\PYZgt{} }\PY{k+kn}{import} \PY{n+nn}{bloscpack} \PY{k+kn}{as} \PY{n+nn}{bp}
\PY{g+gp}{\PYZgt{}\PYZgt{}\PYZgt{} }\PY{n}{a} \PY{o}{=} \PY{n}{np}\PY{o}{.}\PY{n}{linspace}\PY{p}{(}\PY{l+m+mi}{0}\PY{p}{,} \PY{l+m+mi}{100}\PY{p}{,} \PY{l+m+mf}{2e8}\PY{p}{)}
\PY{g+gp}{\PYZgt{}\PYZgt{}\PYZgt{} }\PY{n}{bp}\PY{o}{.}\PY{n}{pack\PYZus{}ndarray\PYZus{}file}\PY{p}{(}\PY{n}{a}\PY{p}{,} \PY{l+s}{\PYZsq{}}\PY{l+s}{a.blp}\PY{l+s}{\PYZsq{}}\PY{p}{)}
\PY{g+gp}{\PYZgt{}\PYZgt{}\PYZgt{} }\PY{n}{b} \PY{o}{=} \PY{n}{bp}\PY{o}{.}\PY{n}{unpack\PYZus{}ndarray\PYZus{}file}\PY{p}{(}\PY{l+s}{\PYZsq{}}\PY{l+s}{a.blp}\PY{l+s}{\PYZsq{}}\PY{p}{)}
\PY{g+gp}{\PYZgt{}\PYZgt{}\PYZgt{} }\PY{k}{assert} \PY{p}{(}\PY{n}{a} \PY{o}{==} \PY{n}{b}\PY{p}{)}\PY{o}{.}\PY{n}{all}\PY{p}{(}\PY{p}{)}
\end{Verbatim}
And here is an example of serializing it to a string:\begin{Verbatim}[commandchars=\\\{\},fontsize=\footnotesize]
\PY{g+gp}{\PYZgt{}\PYZgt{}\PYZgt{} }\PY{k+kn}{import} \PY{n+nn}{numpy} \PY{k+kn}{as} \PY{n+nn}{np}
\PY{g+gp}{\PYZgt{}\PYZgt{}\PYZgt{} }\PY{k+kn}{import} \PY{n+nn}{bloscpack} \PY{k+kn}{as} \PY{n+nn}{bp}
\PY{g+gp}{\PYZgt{}\PYZgt{}\PYZgt{} }\PY{n}{a} \PY{o}{=} \PY{n}{np}\PY{o}{.}\PY{n}{linspace}\PY{p}{(}\PY{l+m+mi}{0}\PY{p}{,} \PY{l+m+mi}{100}\PY{p}{,} \PY{l+m+mf}{2e8}\PY{p}{)}
\PY{g+gp}{\PYZgt{}\PYZgt{}\PYZgt{} }\PY{n}{b} \PY{o}{=} \PY{n}{bp}\PY{o}{.}\PY{n}{pack\PYZus{}ndarray\PYZus{}str}\PY{p}{(}\PY{n}{a}\PY{p}{)}
\PY{g+gp}{\PYZgt{}\PYZgt{}\PYZgt{} }\PY{n}{c} \PY{o}{=} \PY{n}{bp}\PY{o}{.}\PY{n}{unpack\PYZus{}ndarray\PYZus{}str}\PY{p}{(}\PY{n}{b}\PY{p}{)}
\PY{g+gp}{\PYZgt{}\PYZgt{}\PYZgt{} }\PY{k}{assert} \PY{p}{(}\PY{n}{a} \PY{o}{==} \PY{n}{c}\PY{p}{)}\PY{o}{.}\PY{n}{all}\PY{p}{(}\PY{p}{)}
\end{Verbatim}
The compression parameters can be configured as keyword arguments to the
\texttt{pack} functions (see the documentation for detail).

\section{Comparison to NPY%
  \label{comparison-to-npy}%
}

The \cite{NPY} specification lists a number of requirements for the NPY format. To
compare NPY and Bloscpack feature-wise, let us look at the extent to which
Bloscpack satisfies these requirements when dealing with Numpy ndarrays.\newcounter{listcnt0}
\begin{list}{\arabic{listcnt0}.}
{
\usecounter{listcnt0}
\setlength{\rightmargin}{\leftmargin}
}

\item 

\emph{Represent all NumPy arrays including nested record arrays and object arrays.}

Since the support for Numpy ndarrays is very fresh only some
empirical results using toy arrays have been tested. Simple
integer, floating point types and string arrays seem to work fine.
Structured arrays are also supported (as of 0.4.1), even those with
nested data types.  Finally, object arrays also seem to survive the
round-trip tests.
\item 

\emph{Represent the data in its native binary form.}

Since Bloscpack will compress the data it is impossible to represent the data
in its native binary form.
\item 

\emph{Be contained in a single file.}

Using the metadata section of the Bloscpack format all required metadata for
decompressing a Numpy ndarray can be included alongside the compressed data.
\item 

\emph{Support Fortran-contiguous arrays directly.}

If an array has Fortran ordering we can save it in Fortran ordering in
Bloscpack. The order is saved as part of the metadata and the contiguous
memory block is saved as is. The order is set during decompression and hence
the array is deserialized correctly.
\item 

\emph{Store all of the necessary information to reconstruct the array including
shape and dtype on a machine of a different architecture {[}...{]} Endianness
{[}...{]} Type.}

As mentioned above all integer types as well as string  and object arrays are
handled correctly and their shape is preserved. As for endianness, initial
toy examples with large-endian dtypes pass the roundtrip test
\item 

\emph{Be reverse engineered.}

In this case \emph{reverse engineering} refers to the ability to decode a
Bloscpack compressed file after both the Bloscpack code and file-format
specification have been lost. For NPY this can be achieved if one roughly
knows what to look for, namely three metadata attributes and one plain data
block. In the Bloscpack case, things are more difficult. First of all, the
header does have a larger number of entries which must first be deciphered.
Secondly the data is compressed and without knowledge of the compression
scheme and implementation this will be very difficult to reverse engineer.
\item 

\emph{Allow memory-mapping of the data.}

Since the data is compressed it is not possible to use the \DUroletitlereference{mmap}
primitive to map the file into memory in a meaningful way.
However, due to the chunk-wise nature of the storage, it is
theoretically possible to implement a quasi-mem-mapping scheme.
Using the chunk offsets and the typesize and shape from the Numpy
ndarray metadata, it will be possible to determine which chunk or
chunks contain a single element or a range and thus load and
decompress only those chunks from disk.
\item 

\emph{Be read from a file-like stream object instead of an actual file.}

This has been part of the Bloscpack code base since very early versions
since it is essential for unit testing w/o touching the file system, e.g.
by using a file-like \texttt{StringIO} object. In fact this is how the Numpy
ndarray serialization/de-serialization to/from strings is implemented.
\item 

\emph{Be read and written using APIs provided in the numpy package.}

Bloscpack does not explicitly aspire to being part of Numpy.\end{list}

\section{Benchmarks%
  \label{benchmarks}%
}

The benchmarks were designed to compare the following three alternative serialization
formats for Numpy ndarrays: NPY, NPZ and ZFile with Bloscpack. To this end, we
measured compression speed, decompression speed, both with and without the Linux
file system cache and compression ratio for a number of different experimental
parameters.

\subsection{Parameters%
  \label{parameters}%
}

Three different array sizes were chosen:%
\begin{itemize}

\item 

\textbf{small} 1e4 8 = 80000 Bytes = 80KB
\item 

\textbf{mid} 1e7 8 = 80000000 Bytes = 80MB
\item 

\textbf{large} 2e8 * 8 = 1600000000 Bytes = 1.4 GB
\end{itemize}

Three different dataset complexities were chosen:%
\begin{itemize}

\item 

\textbf{low} \texttt{arange} (very low Kolmogorov complexity\DUfootnotemark{id25}{id26}{*})
\item 

\textbf{medium} \texttt{sin} + noise
\item 

\textbf{high} random numbers
\end{itemize}

And lastly two different storage mediums were chosen:%
\begin{itemize}

\item 

\textbf{ssd} encrypted (LUKS) SSD
\item 

\textbf{sd} SD card
\end{itemize}

The SD card was chosen to represent a class of very slow storage, not because
we actually expect to serialize anything to an SD card in practice.

To cut down on the number of data points we choose only to evaluate the
compression levels 1, 3 and 7 for ZFile and 1, 3, 7 and 9 for Bloscpack.
Although NPZ is a compressed format it does not support modifying the
compression level. This results in using \texttt{1 + 1 + 3 + 4 = 9} different
\texttt{codec} values.

This configuration leads to \texttt{3 * 3 * 2 * 9 = 160} data points. Additionally
to account for fluctuations, each datapoint was run multiple times depending on
the size of the dataset. In each case of number of sets each with a number of
runs were performed. Then, the mean across runs for each set and then the
minimum across all sets was taken as the final value for the datapoint. For the
\DUroletitlereference{small} size, 10 sets with 10 runs were performed. For the \DUroletitlereference{mid} size, 5 sets
with 5 runs were performed. And finally, for the \DUroletitlereference{large} size, 3 sets with 3
runs each were performed.%
\DUfootnotetext{id26}{id25}{*}{
The inquisitive reader will note the following caveat at this stage. Perhaps
Kolmogorov complexity is not the correct choice of complexity measure
to define low entropy data for a Lempel-Ziv style dictionary encoder. In fact,
any sequence of consecutive integers by definition has high Lempel-Ziv
complexity and is not compressible. However, as will be shown during the
benchmarks later on, Bloscpack is actually very good at compressing these kinds
of sequences, whereas ZFile and NPZ are not. This is a result of the fact that
\DUroletitlereference{arange} generated muti-byte type integer data and the shuffle filter for
Bloscpack can optimize this very well. At this stage we simply state that the
proposed \textbf{low} entropy dataset has been sufficient for the benchmarks. An
in-depth treatment of the effects the shuffle filter has on the Lempel-Ziv
complexity is beyond the scope of this paper and will perhaps be the subject of
a future publication.}

\subsection{Timing%
  \label{timing}%
}

The timing algorithm used was a modified version of the \texttt{timeit} tool which
included in the Python standard library. This supports deactivation of the
Python interpreters garbage collector during the run and executing code before
and after each run. For example, when measuring decompression speed without the
Linux file system cache, one needs to clear this cache before each run and it
is imperative that this operation does not enter into the timing. Also, when
measuring compression speed, one needs to make sure \texttt{sync} is executed after
the run, to ensure the data is actually written out to the storage medium.
Contrary to clearing the file system cache, the time required by the \texttt{sync}
operation must enter the timing to not contaminate the results.

\subsection{Hardware%
  \label{hardware}%
}

The machine used was a Lenovo Carbon X1 ultrabook with an Intel Core i7-3667U
Processor \cite{CPU}.  This processor has 2 physical cores with active
hyperthreading resulting in 4 threads. The CPU scaling governor was set to
\DUroletitlereference{performance} which resulted in a CPU frequency of 2.0Ghz per core. The CPU has
three levels of cache at: \DUroletitlereference{32K}, \DUroletitlereference{256K} and \DUroletitlereference{4096k} as reported by Linux sysfs.
The memory bandwidth was reported to be 10G/s write and 6G/s read by the Blosc
benchmarking tool.  Interestingly this is in stark contrast to the reported
maximum memory bandwidth of 25G/s which is advertised on the manufacturers data
sheet. The operating system used was Debian Stable 7.1 with the following
64bit kernel installed from Debian Backports:
\DUroletitlereference{3.9-0.bpo.1-amd64 \#1 SMP Debian 3.9.6-1\textasciitilde{}bpo70+1 x86\_64 GNU/Linux}.

The IO bandwidth of the two storage media was benchmarked using \DUroletitlereference{dd}:\begin{Verbatim}[commandchars=\\\{\},fontsize=\footnotesize]
\PY{g+gp}{\PYZdl{}} dd \PY{k}{if}\PY{o}{=}/dev/zero \PY{n+nv}{of}\PY{o}{=}outputfile \PY{n+nv}{bs}\PY{o}{=}512 \PY{n+nv}{count}\PY{o}{=}32M
\PY{g+gp}{\PYZdl{}} dd \PY{k}{if}\PY{o}{=}outputfile \PY{n+nv}{of}\PY{o}{=}/dev/null
\end{Verbatim}
\begin{itemize}

\item SSD: 230 MB/s write / 350 MB/sd read
\item 

SD: 20 MB/sd read/write
\end{itemize}

\subsection{Disabled OS Defaults%
  \label{disabled-os-defaults}%
}

Additionally certain features of the operating system were disabled explicitly
while running the benchmarks. These optimizations were chosen based on empirical
observations while running initial benchmarks, observing suspicious behaviour
and investigating possible causes. While there may be other operating system
effects, the precautions listed next were found to have observably detrimental
effects and disabling them lead to increased reliability of the results.

First, the daily cronjobs were disabled by commenting out the corresponding line
in \texttt{/etc/crontab}. This is important because when running the benchmarks over
night, certain IO intensive cronjobs might contaminate the benchmarks.
Secondly, the Laptop Mode Tools were disabled via a setting in
\texttt{/etc/laptop-mode/laptop-mode.conf}.  These tools will regulate certain
resource settings, in particular disk write-back latency and CPU frequency
scaling governor, when certain system aspects—e.g. the connectivity to AC
power—change and again this might contaminate the benchmarks.

\section{Versions Used%
  \label{versions-used}%
}

The following versions and git-hashes—where available—were used to acquire
the data reported in this article:%
\begin{itemize}

\item 

benchmark-script: NA / 7562c6d
\item 

bloscpack: 0.4.0 / 6a984cc
\item 

joblib: 0.7.1 / 0cfdb88
\item 

numpy: 1.7.1 / NA
\item 

conda: 1.8.1 / NA
\item 

python: 'Python 2.7.5 :: Anaconda 1.6.1 (64-bit)'
\end{itemize}

The benchmark-script and results files are available from the repository of
the  EuroScipy2013 talk about Bloscpack \cite{Haenel2013}. The results file analysed
are contained in the csv file \DUroletitlereference{results\_1379809287.csv}.

\subsection{Bloscpack Settings%
  \label{bloscpack-settings}%
}

In order to reduce the overhead when running Bloscpack some optional features
have not be enabled during the benchmarks. In particular, no checksum is used
on the compressed chunks and no offsets to the chunks are stored.

\section{Results%
  \label{results}%
}

The results of the benchmark are presented in the figures 1, 2, 3, 4 and 5.
Figures 1 to 4 show timing results and are each a collection of subplots where
each subplot shows the timing results for a given combination of dataset size
and entropy. The dataset size increases horizontally across subplots whereas
the dataset entropy increases vertically across subplots. Figures 1 and 2 show
results for the SSD storage type and figures 3 and four show results for the SD
storage type. Figures 1 and 3 compare Bloscpack with NPY whereas figures 2 and
4 compare Bloscpack with NPZ and ZFile. NPY is shown separately from NPZ and
ZFile since their performance characteristics are so different that they can not
be adequately compared visually on the same plot. For all timing plots black
bars indicate compression time, white is used to denote decompression time w/o
the file system cache and gray identifies decompression time with a hot file system
cache. For all timing plots, larger values indicate worse performance. Lastly,
figure 5 shows the compression ratios for all examined formats.\begin{figure*}[figclass]\noindent\makebox[\textwidth][c]{\includegraphics[scale=0.60]{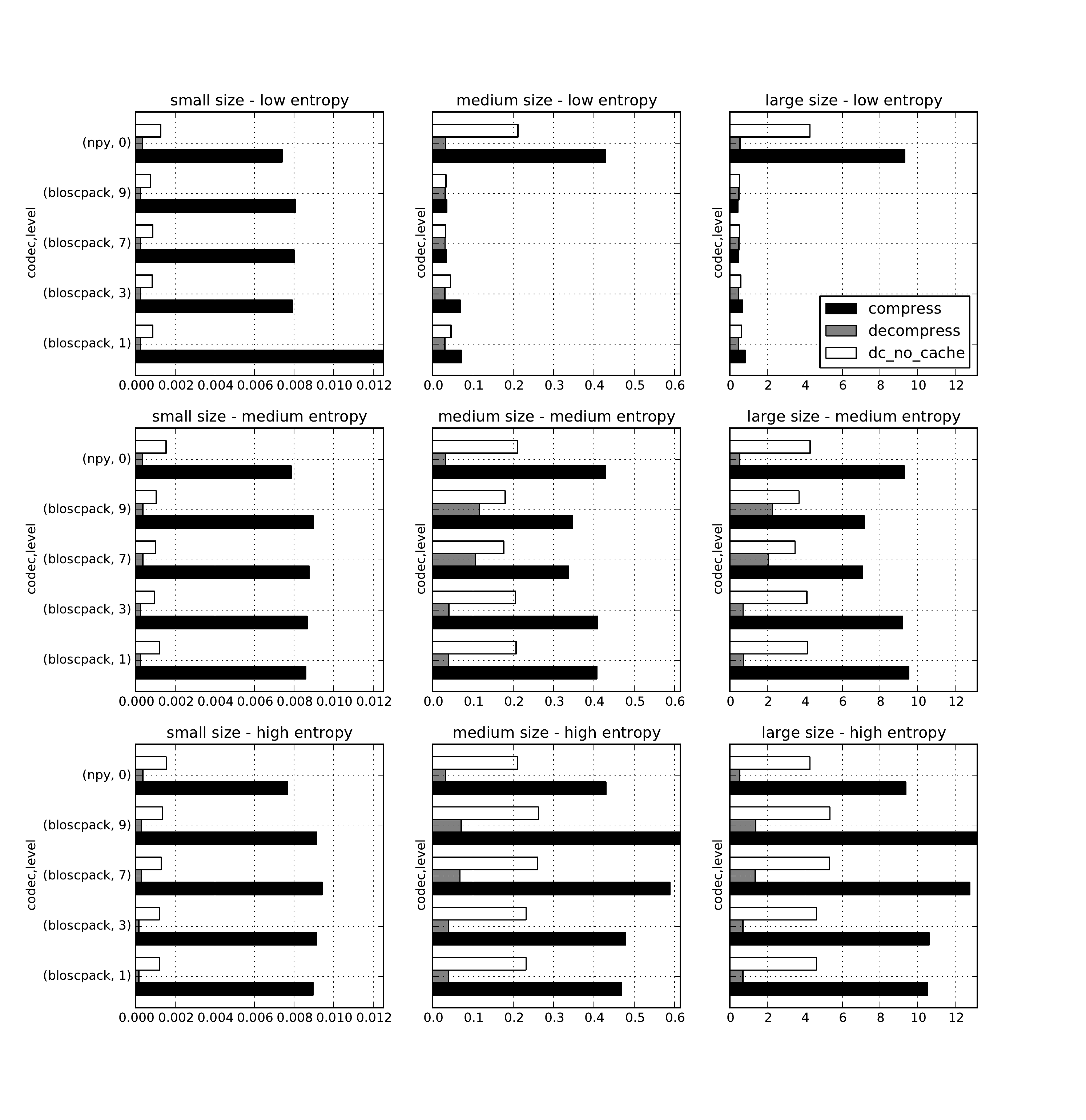}}
\caption{Compare Bloscpack and NPY on the SSD storage type.}
\end{figure*}\begin{figure*}[figclass]\noindent\makebox[\textwidth][c]{\includegraphics[scale=0.60]{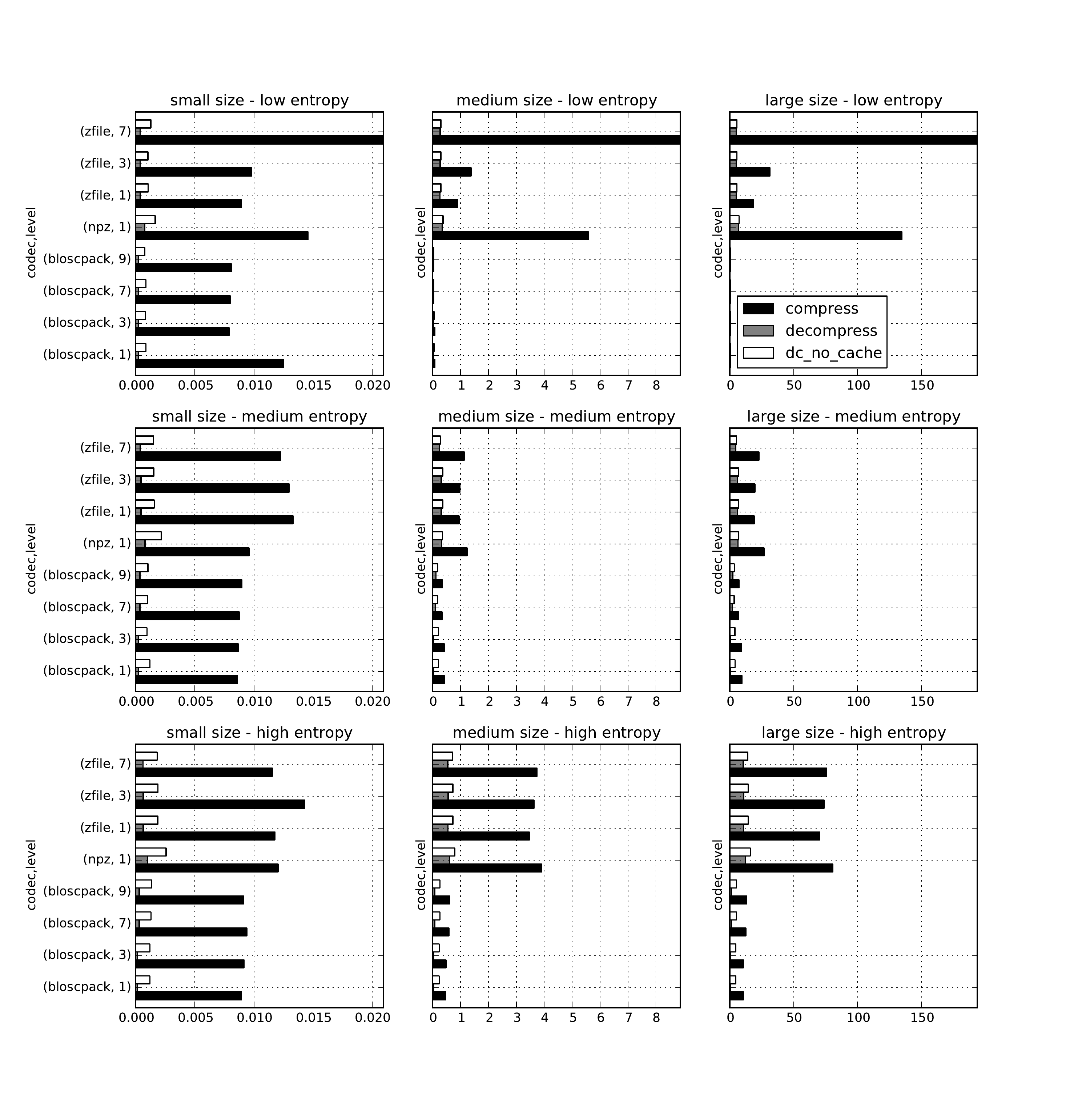}}
\caption{Compare Bloscpack, NPZ and ZFile on the SSD storage type.}
\end{figure*}\begin{figure*}[figclass]\noindent\makebox[\textwidth][c]{\includegraphics[scale=0.60]{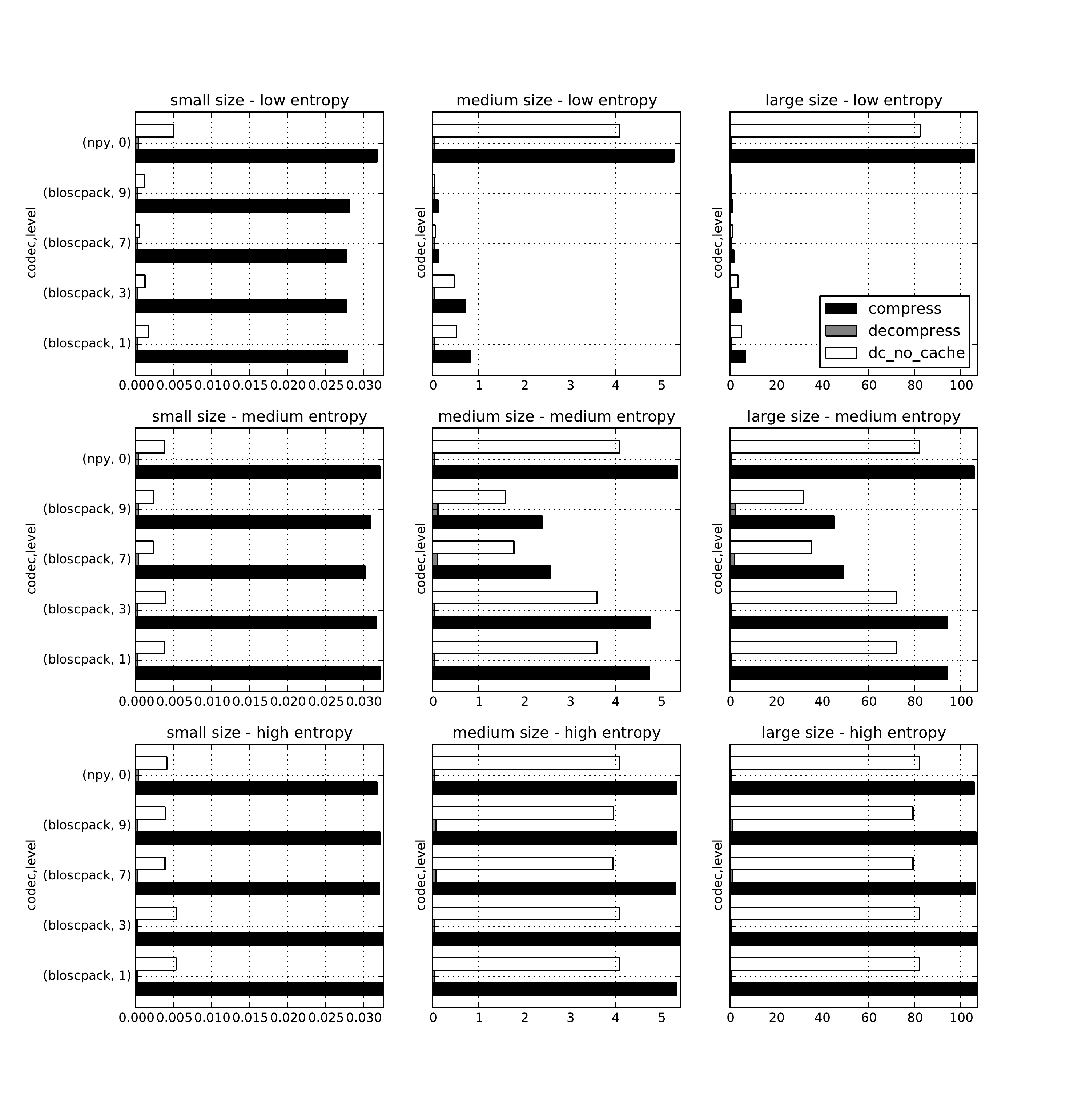}}
\caption{Compare Bloscpack and NPY on the SD storage type.}
\end{figure*}\begin{figure*}[figclass]\noindent\makebox[\textwidth][c]{\includegraphics[scale=0.60]{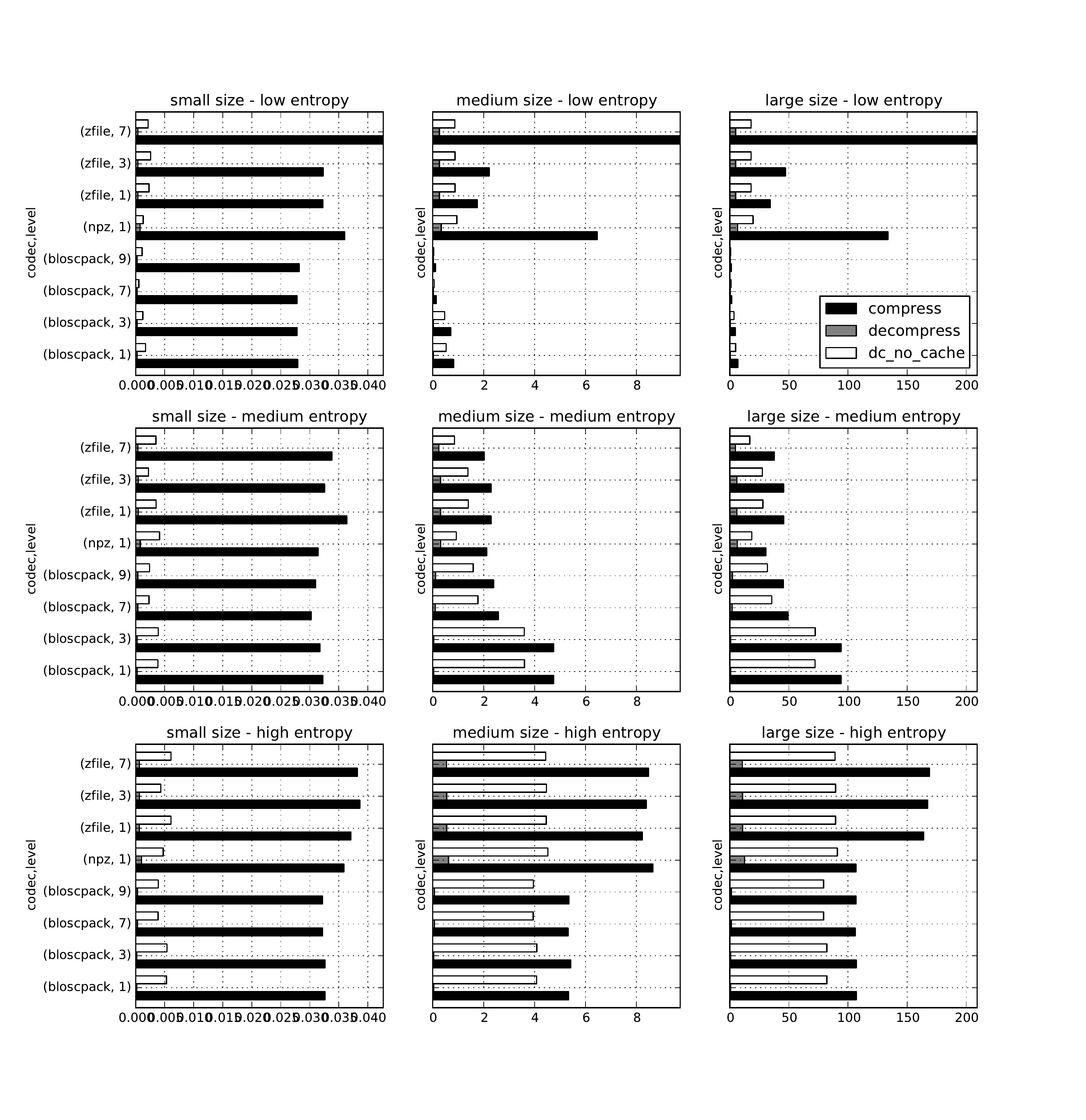}}
\caption{Compare Bloscpack, NPZ and ZFile on the SD storage type.}
\end{figure*}\begin{figure*}[figclass]\noindent\makebox[\textwidth][c]{\includegraphics[scale=0.60]{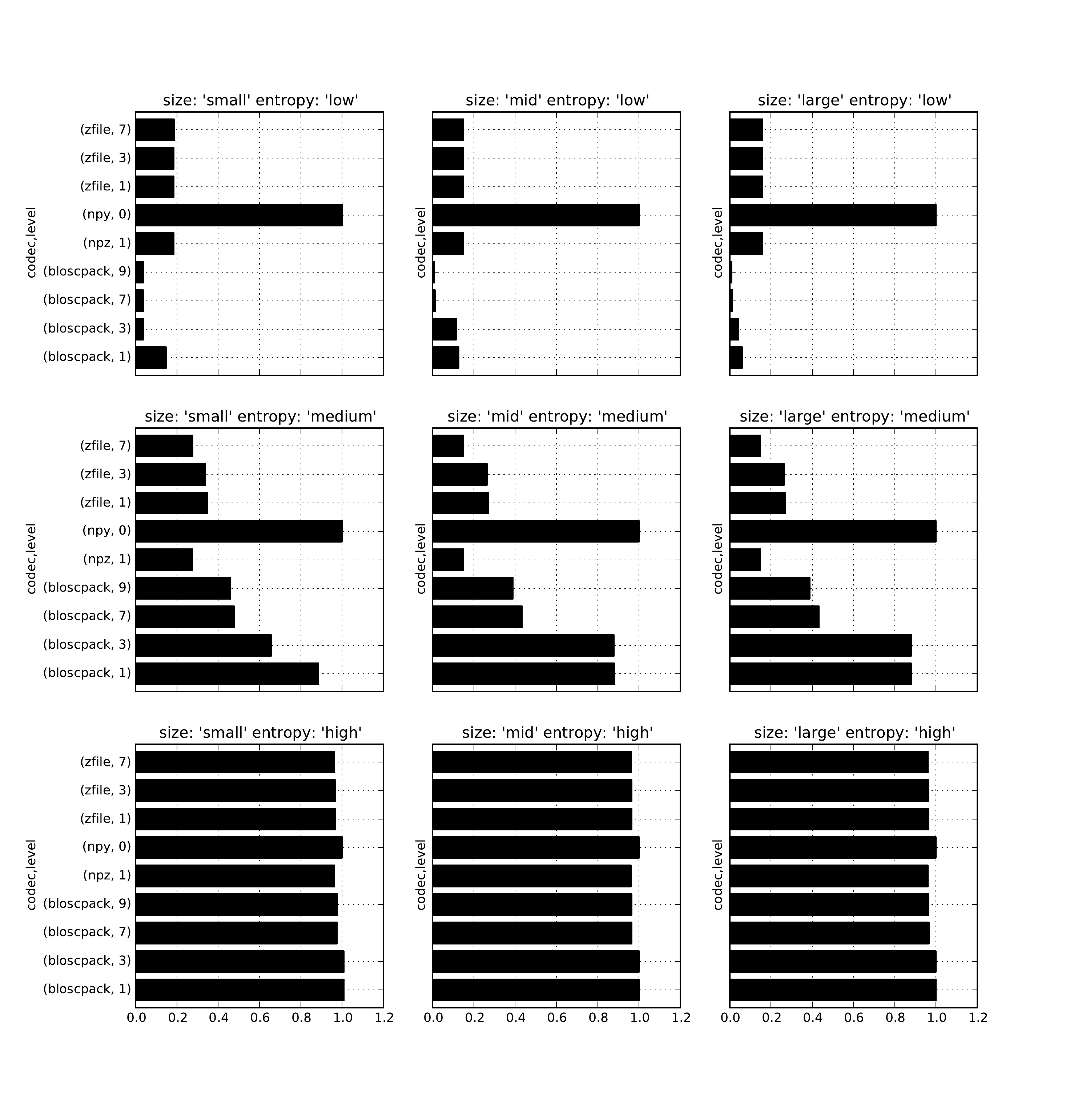}}
\caption{Compression ratios for all examined formats}
\end{figure*}

In Fig. 1 we can see how Bloscpack compares to NPY on the SSD storage type. The
first thing to note, is that for small datasets (first column of subplots),
Bloscpack does not lag behind much compared to NPY for compression and is
actually slightly faster for decompression. However the absolute differences
here are in the millisecond range, so one might perhaps argue that Bloscpack and
NPY are on par for small datasets. As soon as we move to the medium size
datasets first gains can be seen. Especially for the low entropy case where
Bloscpack beats NPY for both compression and decompression w/o file system
cache. For the medium entropy case, Bloscpack is slightly faster for a few
settings, at least for the compression and decompression cases. Surprisingly,
for the decompression with a hot file system cache, Bloscpack is actually 2
times slower under the compression levels 7 and 9. One possibility for this
might be that, even though the file contents are in memory, reading from the
file necessitates an initial memory-to-memory copy, before the data can
actually be decompressed.  For the high entropy case, Bloscpack is mostly
slightly slower than NPY. For the large dataset the results are simply a scaled
version of the medium dataset size results and yield no additional insights.

Fig. 2 shows the comparison between Bloscpack, NPZ and ZFile on the SSD storage
type. In this comparison, the speed of the Blosc compressor really shines. For
every combination of dataset size and entropy the is a compression level for
Bloscpack that can compress faster than any of the competitors. In the extreme
case of the large size and the low entropy, Bloscpack is over 300 times faster
during compression than NPZ (302 seconds for NPZ vs. 0.446 seconds for
Bloscpack).  Even for the high entropy case, where very very little compression
is possible due to the statistics of the dataset, Bloscpack is significantly
faster during compression.  This is presumably because Blosc will try to
compress a buffer, finish very quickly because there is no work to be done and
then it simply copies the input verbatim.

One very surprising result here is that both NPZ and ZFile with level 7 take
extraordinary amounts of time to compress the low entropy dataset. In fact they
take the longest on the low entropy dataset compared to the medium and high
entropies. Potentially this is related to the high Lempel-Ziv complexity of
that dataset, as mentioned before. Recall that both NPZ and ZFile use the
DEFLATE algorithm which belongs to the class of LZ77 dictionary encoders, so it
may suffer since it no shuffle filter as in the case of Blosc is employed.

Figures 3. and 4. show the same results as figures 1. and 2. respectively but
but for the SD storage class. Since the SD card is much slower than the SSD
card the task is strongly IO bound and therefore benefits of compression can be
reaped earlier. For example, Bloscpack level 7 is twice as fast as NPY during
compression on the medium size medium entropy dataset. For the low entropy
dataset at medium and large sizes, Bloscpack is about an order of magnitude
faster.  For the high entropy dataset Bloscpack is on par with NPY because the
overhead of trying to compress but not succeeding is negligible due to the IO
boundedness resulting from the speed of the SD card. When comparing Bloscpack
to NPZ and ZFile on the SD card, the IO boundedness means that any algorithm
that can achieve a high compression ratio in a reasonable amount of time will
perform better. For example for medium size and medium entropy, NPZ is actually
1.6 times faster than Bloscpack during compression. As in the SSD case,
we observe that NPZ and ZFile perform very slowly on low entropy data.

Lastly in Figure 5. we can see the compression ratios for each codec, size and
entropy. This is mostly just a sanity check. NPY is always at 1, since it is a
plain serialization format. Bloscpack gives better compression ratios for low
entropy data. NPZ and ZFile give better compression ratios for the medium
entropy data. And all serializers give a ratio close to zero for the high
entropy dataset.

\section{Conclusion%
  \label{conclusion}%
}

This article introduced the Bloscpack file-format and python reference
implementation. The features of the file format were presented and compared to
other serialization formats in the context of Numpy ndarrays. Benchmarking
results are presented that show how Bloscpack can yield performance
improvements for serializing Numpy arrays when compared to existing solutions
under a variety of different circumstances.

\section{Future Work%
  \label{future-work}%
}

As for the results obtained so far, some open questions remain unsolved. First
of all, it is not clear why Bloscpack at level 7 and 9 gives comparatively bad
results when decompressing with a hot file system cache. Also the bad
performance of ZFile and NPY on the so-called low entropy dataset must be
investigated and perhaps an alternative can be found that is not biased towards
Bloscpack.  Additionally, some mathematical insights into the complexity reduction
properties of Blosc's shuffle filter would be most valuable.

Lastly, more comprehensive benchmarks need to be run. This means, first finding
non-artificial benchmark datasets and establishing a corpus to run Bloscpack
and the other solutions on. Furthermore, It would be nice to run benchmarks on other
architectures for machines with more than 2 physical cores, non-uniform memory
access and an NFS file-system as commonly found in compute clusters.

\section{Gratitude%
  \label{gratitude}%
}

The author would like to thank the following people for advice, helpful
comments and discussions: Pauli Virtanen, Gaël Varoquaux, Robert Kern and
Philippe Gervais. Also, the author would like to specially thank Stéfan van der
Walt and Francecs Alted for reviewing drafts of this paper.

\end{document}